\documentclass[floatfix,aps,twocolumn,nofootinbib,superscriptaddress,longbibliography]{revtex4-2}

\usepackage{siunitx}

\usepackage{graphicx}
\usepackage{epsfig}

\usepackage{amsmath}
\usepackage{amsfonts}
\usepackage{amssymb}
\usepackage{bm}
\usepackage{mathtools}

\usepackage{dsfont}

\usepackage{tikz}
\usepackage{circuitikz}
\usetikzlibrary{arrows, shapes}
\usepackage{pgfplots}
\pgfplotsset{compat=1.14}

\usepackage{macros}

\newcommand{\mean}[1]{\left\langle #1 \right\rangle}

\usepackage{listings}
\usepackage{xcolor}

\definecolor{dark-red}{rgb}{0.4,0.15,0.15}
\definecolor{dark-blue}{rgb}{0.15,0.15,0.4}
\definecolor{medium-blue}{rgb}{0,0,0.5}
\usepackage[hypertexnames=false]{hyperref}
\usepackage[all]{hypcap}
\hypersetup{
	colorlinks, linkcolor={dark-red},
	citecolor={dark-blue}, urlcolor={medium-blue}
}

\definecolor{bkgd}{RGB}{240,242,246}
\definecolor{ceruleanblue}{rgb}{0.16, 0.32, 0.75}
\definecolor{orange-red}{rgb}{1.0, 0.27, 0.0}
\definecolor{anotherblue}{RGB}{37,92,243}
\definecolor{blackblue}{RGB}{46,60,85}
\definecolor{goldyellow}{RGB}{199,146,12}

\newcommand{\revcolor}{black}

\lstdefinestyle{altstyle2}{
	backgroundcolor=\color{bkgd},
	basicstyle=\ttfamily\footnotesize\color{blackblue},
	breakatwhitespace=false,
	breaklines=true,
	captionpos=b,
	commentstyle=\color{goldyellow},
	keepspaces=true,
	keywordstyle=\color{orange-red},
	language=Python,
	numbersep=5pt,
	numberstyle=\tiny\color{ceruleanblue},
	showspaces=false,
	showstringspaces=false,
	showtabs=false,
	stringstyle=\color{anotherblue},
	tabsize=2
}
\usepackage{hyperref}
\usepackage{dsfont}

\begin{document}
\title{Taming non-equilibrium thermal fluctuations in subthreshold CMOS circuits}

\author{Nahuel Freitas}\thanks{These authors contributed equally to this work.}
\author{Geremia Massarelli}
\author{Jeremy Rothschild}
\author{Dylan Keane}
\author{Ethan Dawe}
\author{Sewook Hwang}
\author{Akhil Garlapati}
\thanks{These authors contributed equally to this work.}

\author{Trevor McCourt$\,^{*,\,}$}
\email[Corresponding authors:\hspace{0.3em}]{nfreitas@extropic.ai, trevor@extropic.ai}

\affiliation{Extropic Corporation, Cambridge, Massachusetts, USA}

\date{\today}

\begin{abstract}
As CMOS technology scales down, thermal fluctuations increasingly impact circuit behavior, posing challenges to conventional circuit design. However, the inherent stochasticity introduced by thermal noise is now being explored as a potential resource in the emerging field of probabilistic computing. This work presents a fully CMOS experimental platform that enables direct control over its intrinsic thermal fluctuations. These devices function as programmable multivariate Gaussian samplers, offering a hardware primitive for energy-efficient stochastic computing and serving as an experimental platform for studies in electronic noise and stochastic thermodynamics.
\end{abstract}

\maketitle

\begin{figure*}
    \centering
    \includegraphics[width=1.0\linewidth]{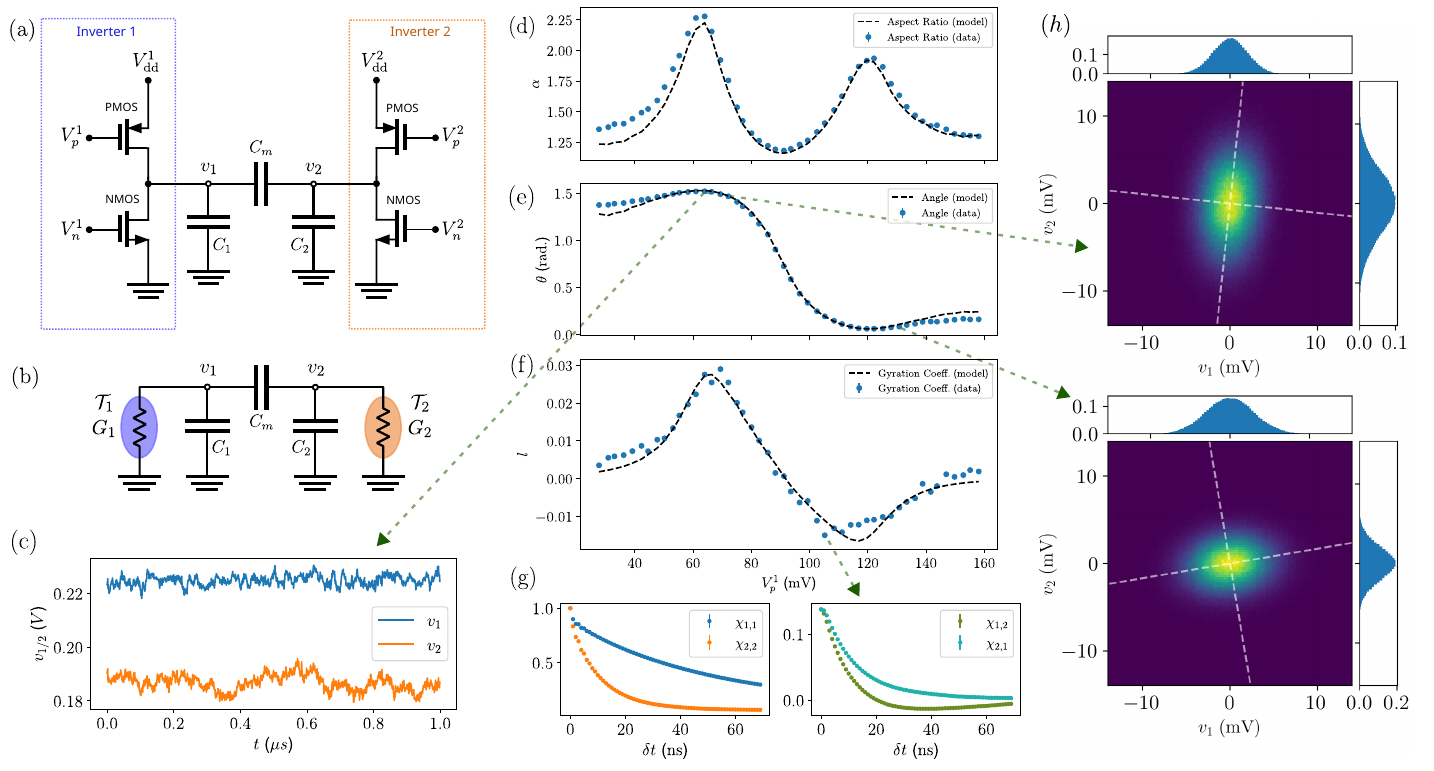}
    \caption{\textbf{Predictable and programmable Gaussian sampling with a NEAT-RN (a)} A simple NEAT-RN: two independently controlled, capacitively coupled CMOS inverters. \textbf{(b)} The fluctuations around the DC operating point of a NEAT-RN are equivalent to those of a linear RC circuit with controllable effective conductances and temperatures. \textbf{(c)} Snapshot of the measured time-series of voltages $v_1$ and $v_2$, for the operating point indicated by the arrow. For different points along the voltage sweep mentioned in the text: \textbf{(d)} Aspect ratio of the covariance matrix of signals $v_1$ and $v_2$, \textbf{(e)} Angle of the main principal component of the same matrix, and \textbf{(f)} Normalized gyration coefficient $l$ (integration time $t_I = 200\unit{ns}$).  \textbf{(g)} Example of the delayed autocorrelation (left) and crosscorrelation (right) functions. \textbf{(h)} Example centered histograms of the signals $v_1$ and $v_2$ for two operating points. Dashed lines indicate the principal axes. In all cases we fixed the powering voltages $V_\text{dd}^{1} = 137\pm 1 \unit{\milli\volt}$, $V_\text{dd}^{2} = 142 \pm 1\unit{\milli\volt}$ and NMOS gate voltages $V_\text{n}^{1}=V_\text{n}^{2}=76 \pm 1  \unit{\milli\volt}$, while PMOS gate voltages satisfy $V_p^1 + V_p^2 = 200 \pm 1\unit{\milli\volt}$.}
    \label{fig:main_figure}
\end{figure*}

The aggressive miniaturization of CMOS technology is a double-edged sword; as a transistor circuit is scaled down in size, it becomes faster and more energy-efficient \cite{dennard}, but it also becomes noisier \cite{nyquist_thermal}. As device dimensions and supply voltages continue to be pushed to their thermodynamic limits, the noise that is intrinsic to transistors will begin to upset traditional deterministic computer architectures \cite{kish2002end, rezaei2020fundamental}.

Recent proposals suggest that this apparent roadblock imposed by noise may be circumvented by building intentionally stochastic \textit{probabilistic circuits} (p-circuits) into a computing system \cite{zink2022review,chowdhury2023full,kim2024fully,ahmed2021probabilistic, Lee2025Mar, Kim2018Feb, Borders2019Sep}. P-circuits harness the noise present in all devices to generate samples from computationally useful probability distributions.

Compared to purely deterministic approaches, probabilistic computer architectures can be more energy-efficient when running specific algorithms that rely heavily on random sampling. Random sampling is expensive on a deterministic computer: one must employ complex circuitry involving thousands of transistors that generate pseudorandom numbers \cite{singh2024cmos}. These large circuits consume similar amounts of energy to those implementing computationally richer operations, such as addition.

For a p-circuit to be practically useful in computing, it must be predictable, programmable, and performant. Predictability means that the stochastic dynamics of the p-circuit obey some simple physical model that can be used to engineer it. A p-circuit is programmable if the distribution it samples from can be tuned at runtime. The performance of a p-circuit is measured by its correlation time, energy consumption, and physical size, which jointly quantify the space-time-energy cost of using the circuit for random sampling. 

Current implementations of p-circuits fall short in at least one of these key areas. P-circuits built using magnetic tunnel junctions are a promising long-term direction as they obey well-characterized stochastic dynamics \cite{smtj_model} and are straightforwardly programmed \cite{camsari2017stochastic, camsari2017implementing, shao2023probabilistic}. While these devices are, in principle, compatible with CMOS processes \cite{vodenicarevic2017low}, work on integration is still underway \cite{yang201828nm, edelstein202014}, which limits near-term performance \cite{daniel2024experimental, camsari2017situ}. \textcolor{\revcolor}{To avoid these integration issues, practitioners have also considered p-circuit architectures involving only transistors \cite{korkmaz2006advocating, palem2013ten}. However, current approaches generally rely on extremely sensitive phenomena, such as the randomness of bistable latches at power-up \cite{holleman20083, bae202415}, or are based on crude treatments of noise \cite{korkmaz2006advocating}, making programmability and prediction difficult.}

Circuits of transistors operating in the deep subthreshold regime are a natural candidate for building practical p-circuits. Charge transport in these circuits is predominately thermally activated, leading to shot noise dynamics  \cite{sarpeshkar1993white}. Recently, shot noise models have been generalized from the single-transistor level to networks of transistors via a technique based on Markov jump processes (MJPs) \cite{freitas2021stochastic}, suggesting that the stochastic dynamics of subthreshold circuits may be rigorously predicted. Subthreshold networks are also straightforwardly programmable via control voltages. Additionally, small transistors leak more current, increasing the performance of subthreshold circuits \cite{morifuji2006supply}.

This article experimentally demonstrates that practical p-circuits can be built using subthreshold transistors. Specifically, we introduce and characterize a family of circuits that sample from voltage-programmable multivariate Gaussian distributions. These circuits enable the control of correlations between multiple degrees of freedom and represent a powerful primitive that can be used either as a standalone p-circuit or as part of a larger circuit performing a non-Gaussian sampling operation \textcolor{\revcolor}{(p-bits being the most direct and prominent example \cite{Borders2019Sep}}). 
Our work paves the way for near-term, large-scale computers that leverage p-circuits; we have already applied the principles outlined here in designing a probabilistic computing system that runs diffusion-like models \cite{jelincic2025}.

In addition to the potential of our work for probabilistic computing, it also substantially advances our physical understanding of shot noise in subthreshold transistors by validating the MJP model presented in \cite{freitas2021stochastic}. While a large body of work studies shot noise at the single-transistor level \cite{reimbold1982white, tedja1992noise, sarpeshkar1993white}, studies of noise physics in larger circuits are limited and often application-specific \cite{tian1999analysis}. Our work represents the first time a generic and simple noise model for subthreshold transistor networks has been explicitly tested against experimental data, allowing the model to be confidently used in future engineering pursuits.

The remainder of this article will detail our Gaussian sampling circuits, which we refer to as non-equilibrium, adjustable temperature resistor networks (NEAT-RNs). We will first introduce NEAT-RNs and explain (as suggested by their name) how voltage controls can be used to program their steady-state voltage distribution. Then, we will demonstrate that the MJP model of \cite{freitas2021stochastic} accurately predicts experimental measurements of the behavior of a simple 2D NEAT-RN. Next, we will demonstrate that the programmability of NEAT-RNs can be significantly enhanced by introducing additional degrees of freedom that mediate interaction. Finally, we will discuss the speed and energy efficiency of NEAT-RNs and demonstrate that they are sufficiently performant to be useful as part of a modern probabilistic computing system.

NEAT-RNs are composed of multiple CMOS inverters with independent gate control, where mutual capacitances couple the outputs of different inverters. The simplest non-trivial example in the family is shown in Fig. \ref{fig:main_figure}-(a). \textcolor{\revcolor}{The experiments discussed in the following involve the measurement of the statistical properties of the signals $v=(v_1,v_2)$ for different operating voltages (see Fig \ref{fig:main_figure}-(c) for examples of time series). Most of the variability in those signals is explained by two sources of noise: i) thermal fluctuations associated to the transport of charge through individual transistors, and ii) low-frequency fluctuations in the overall conductivity of each transistor. While the first contribution is of fundamental nature, as it has a purely thermodynamic origin \cite{sarpeshkar1993white, freitas2021stochastic}, the second one is usually understood to arise due to material defects that act as fluctuating charge traps \cite{mahmutoglu2013modeling,stampfer2020advanced,tian2002analysis,jakobson19981,kirton1989noise}. The obtained data is compared to an analytical model that is able to incorporate both sources of noise. However, in what follows we restrict the discussion to the analytical description of the thermal fluctuations only, which are anyway the dominant ones as they contribute around $70\%$ of the total signal power in the frequency range given by the measurement duration and sampling rate (see Appendix \ref{ap:spectral_analysis}).}

After linearization around an operating point, NEAT-RNs become equivalent to a noisy RC network \cite{ciliberto2013statistical, baiesi2016thermal, Freitas2020Jul}, where the effective conductance and temperature of each resistor can be controlled at will (see Fig. \ref{fig:main_figure}-(b)). 
When the effective temperatures are equal, the system attains an equilibrium steady-state probability distribution $P(v)$ that is entirely determined by the capacitances, which cannot be controlled in nanoscopic circuits. However, if the effective temperatures differ, the system attains a non-equilibrium steady-state distribution that becomes dependent on the effective temperatures and conductances. In this way, the steady-state distribution can be controlled by manipulating the voltages $V_\text{dd}^{i}$, and $V_{n/p}^{i}$ (see Fig. \ref{fig:main_figure}-(f-g)).  

Specifically, combining the MJP formalism of \cite{freitas2021stochastic} with the Enz-Krummenacher-Vittoz (EKV) model for subthreshold transistors \cite{Enz2006Jul}, the steady state fluctuations of a NEAT-RN can be found to be approximately Gaussian,
\begin{equation}
    P(v) \propto \exp{ -\frac{1}{2} \left( v - v^* \right)^T\Sigma^{-1} \left( v - v^* \right).}
    \label{eq:gaussian_ss}
\end{equation}
Here, $v$ is a vector of voltages for each node in the circuit, $v^*$ is the DC level of each voltage, and $\Sigma$ is the steady-state covariance matrix. $\Sigma$ is given by the solution to the Lyapunov equation,
\begin{equation}
    G \: \Sigma \: C  + C \: \Sigma \: G = 2k_b \: G^{1/2} \:
    \mathcal{T} \: G^{1/2},
    \label{eq:lyapunov}
\end{equation}
where $G$ is a diagonal matrix of effective linear conductances associated with each inverter, $\mathcal{T}$ is a diagonal matrix of effective temperatures, and $C$ is the circuit's Maxwell capacitance matrix. The matrix elements of $\mathcal{T}$ and $G$ can be written explicitly in terms of the circuit's control voltages; see Appendixes \ref{ap:diffusive_limit} and \ref{ap:ekv_model}.

In the limit that each inverter is balanced such that $v^{*}_i = V_\text{dd}^i/2$ and the subthreshold slope parameter of each transistor approaches 1 (see Appendix C), the expressions for the matrix elements of $G$ and $\mathcal{T}$ take an intuitive form,
\begin{equation}\label{eq:g_exp}
    G_{ii} = 2 \frac{I_0^p}{V_T} \exp\left(\left(V_\text{dd}^i/2 - V_p^i \right)/V_T\right)
\end{equation}
\begin{equation}\label{eq:t_exp}
    \mathcal{T}_{ii} = T \frac{1 + \exp\left(V_\text{dd}^i/2 V_T\right)}{2}
\end{equation}
Here, $I_0^p$ is the EKV model current parameter for each transistor. $T$ is the physical temperature of the system and $V_T=k_bT/q_e$ the associated thermal voltage ($q_e$ is the electron charge). We chose to eliminate the control voltage $V_{n}^{i}$ from this expression using the balance constraint. This decision was arbitrary, and the expressions could be written just as easily with $V_{p}^{i}$ eliminated.

This limiting case makes it clear that NEAT-RNs are highly programmable; the effective conductance of each branch can be set by varying the gate control voltage with respect to $V_\text{dd}$, and the effective temperature can be elevated with respect to the actual ambient temperature by increasing $V_\text{dd}^{i}$ above ground. The potential values that $G_{ii}$ and $\mathcal{T}_{ii}$ can take are limited by the onset of moderate inversion, at which point the shot noise modeling assumption used to derive Eq.~\eqref{eq:lyapunov} becomes invalid. As long as the devices are in subthreshold, $\mathcal{T}$ and $G$ can be independently controlled to program the p-circuit with a wide range of covariance matrices.

To empirically establish the predictability and programmability of our p-circuit, we fabricated the circuit shown in Fig.~\ref{fig:main_figure} (a) using an advanced TSMC FinFET \cite{sekigawa1984calculated, wu20223nm, tsmc5} process and fit Eq.~\eqref{eq:lyapunov} to its steady state voltage distribution. Specifically, we fixed all voltages except for the PMOS control voltages, which were swept such that
$V_p^1 + V_p^2 = 200 \pm 1\unit{\milli\volt}$. Voltage time series were measured at each operating point, from which a covariance matrix was estimated (the specifics of the fitting procedure are outlined in Appendix \ref{ap:data_analysis}).

We characterized the measured covariance matrix via its spectrum. Specifically, we computed the eigenvalues $\lambda_k$ and eigenvectors $q_k$, satisfying $\Sigma q_k = \lambda_k q_k$
with $\lambda_0 \geq \lambda_1$.
From the eigenvalues and eigenvectors of $\Sigma$, we define the aspect ratio of the eigenvalues $\alpha$ and the angle of the principal component $\theta$,
\begin{equation}
    \alpha \equiv \sqrt{\lambda_0/\lambda_1}, \qquad
    \theta \equiv \arctan((q_0)_1/(q_0)_2).
\end{equation}
Examples of the observed steady-state distribution are shown in Figs.~ \ref{fig:main_figure} (f) and (g). 

Figs.~\ref{fig:main_figure} (c) and (d) compare the observed dependence of $\alpha$ and $\theta$ on $V_p^1$ to the best fit of Eq.~\eqref{eq:lyapunov}. These results indicate that the fluctuations in the signals $v_1$ and $v_2$ can be reliably controlled by changing the operating voltages $V_{n/p}^{1/2}$. Additionally, we observe that the simple modeling based on Eq.~\eqref{eq:lyapunov} effectively captures the main features of the data. Thus, the produced signals are both controllable and predictable.

To establish that the MJP model can predict the dynamic properties of a NEAT-RN, not just the steady-state distribution, we measure the circulation of the voltage state and compare it to the prediction of the model. This circulation can be quantified via the coefficient $L(\delta t) \equiv \mean{v \times \delta v}$, where $\delta v = v (t+\delta t) - v(t)$ is the displacement in time $\delta t$. At steady-state conditions, it is possible to rewrite $L$ as
\begin{equation}
    L(\delta t) = \mathbb{X}_{1,2}(\delta t) - \mathbb{X}_{2,1}(\delta t),
    \label{eq:def_L}
\end{equation}
where $\mathbb{X}_{j,k}(\delta t) \equiv \langle v_j(t) v_k(t+\delta t)\rangle - \langle v_j(t) \rangle\langle v_k(t+\delta t) \rangle$ is a delayed correlation, which can be obtained from time series. The previous expression is known as the cross-correlation asymmetry and is a measure of time-reversal symmetry breaking \cite{Ohga2023Aug, Liang2023Dec, Gu2024Apr}.
For example, in Fig. \ref{fig:main_figure}-(g) we show the normalized delayed correlations $\chi_{j,k} \equiv \mathbb{X}_{j,k}/\sqrt{\mathrm{var}(v_j) \mathrm{var}(v_k)}$ for a particular operating point. We observe that the delayed correlations are indeed asymmetric, indicating a breakdown of time-reversal invariance due to the non-equilibrium conditions.

By analyzing the stochastic differential equation that describes the dynamics of a NEAT-RN, it is possible to derive the following expression for time averages of delayed correlations (see Appendix \ref{ap:gyration}):
\textcolor{\revcolor}{
\begin{equation}
\begin{split}
    \bar{\mathbb{X}}_{j,k} &= \frac{1}{t_I} \int_0^{t_I} d\tau \: \mathbb{X}_{j,k}(\tau) \\
    &= \frac{1}{t_I}\left[ \Sigma \left(\mathds{1} - e^{-t_IGC^{-1}}\right) C G^{-1} \right]_{j,k}
    \label{eq:delayed_corr}
\end{split}
\end{equation}
Fig.~\ref{fig:main_figure}-(e) shows the normalized circulation coefficient ${l = (\bar{\mathbb{X}}_{1,2}-\bar{\mathbb{X}}_{2,1})/\sqrt{\mathrm{var}(v_1)\mathrm{var}(v_2)}}$ for an integration time of $t=200 \unit{ns}$ and the comparison with the model best fit. 
}
As before, the agreement shows that the statistical properties of the signals can be reliably controlled and predicted. 
 
Additionally, these results show that the device in Fig. \ref{fig:main_figure}-(a) can be considered the first electronic implementation of a Brownian gyrator, an elementary heat engine \cite{filliger2007brownian}, using CMOS circuits. Previous implementations of Brownian Gyrators have varied the environmental temperature directly by heating or cooling the circuit \cite{chiang2017electrical}, which is more cumbersome than our electronic implementation.

\begin{figure}
    \centering
    \includegraphics[width=0.95\linewidth]{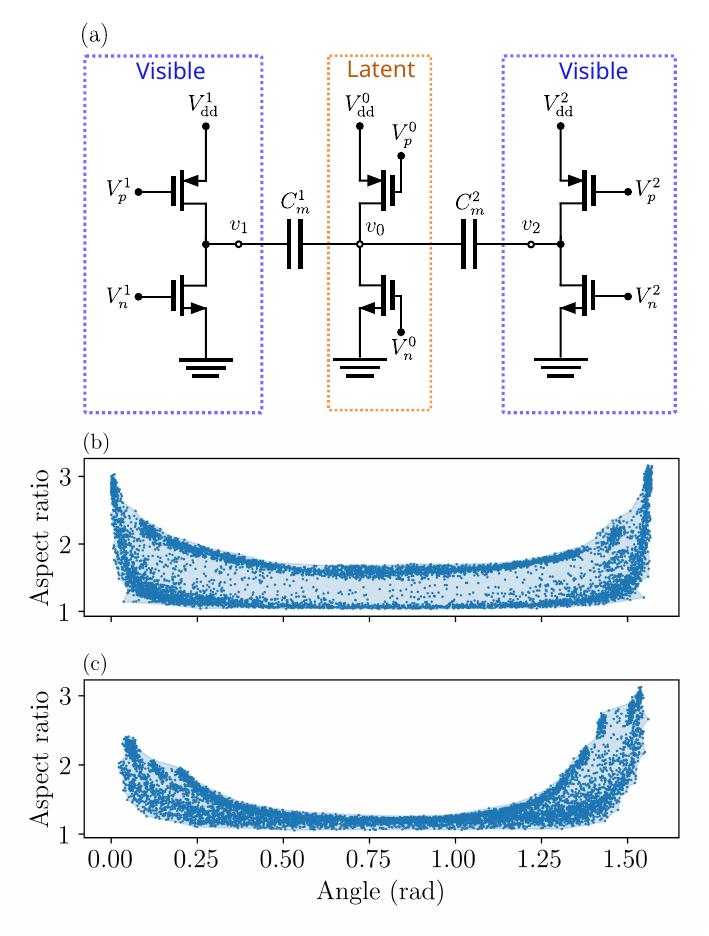}
    \caption{\textbf{Increasing the programmability of a NEAT-RN using latent variables (a)} Extending the circuit in Figure \ref{fig:main_figure}-(a) by adding an extra degree of freedom. The self-capacitance of each free node is omitted. \textbf{(b)} Observed points in the plane angle-aspect ratio for $8000$ iterations of the control space exploration procedure, with $V_\text{dd}^{1/2} \simeq 5.5V_T$ and $V_\text{dd}^0 \simeq 7 V_T$. \textbf{(c)} Results of an analogous protocol applied to the circuit in Figure \ref{fig:main_figure}-(a).}
    \label{fig:3x_gyrator}
\end{figure}

NEAT-RNs with additional nodes allow for richer programmability than the minimal example considered thus far. In such circuits, one can split the nodes into visible and latent nodes. Then, the correlations among visible nodes can be shaped by their common interaction with latent ones, which naturally enhances the range of achievable distributions.

To experimentally establish this enhanced programmability, we built a three-degree-of-freedom (DOF) NEAT-RN that utilizes a latent inverter to cover a larger range of 2D distributions, as shown in Fig.~\ref{fig:3x_gyrator} (a). The additional DOF mediates the coupling between the two outputs, allowing for richer control of their correlation.

We studied the programmability of our NEAT-RNs by exploring their control spaces for extreme covariance matrices. Specifically, for fixed powering voltages $V_\text{dd}^{0/1/2}$, we generated gate voltages $V_{n/p}^{i}$ (in the range $[0,1.75\unit{\volt}]$) via a sampling procedure designed to find configurations with extreme aspect ratios. The results are shown in Fig. \ref{fig:3x_gyrator}.

By comparing Figs.~\ref{fig:3x_gyrator}-(b) and (c), we can see that our 3-DOF NEAT-RN is much more programmable than the 2-DOF version. For any given value of $\theta$, the 3-DOF circuit allows $\alpha$ to be controlled within a much larger range than the 2-DOF circuit. In fact, for values of $\theta$ near $\pi/4$, $\alpha$ can barely be controlled for the 2-DOF circuit, whereas the 3-DOF circuit still features full programmability. 

For a circuit to be an efficient entropy source in a probabilistic computer, it must generate fluctuations large enough to affect downstream devices. For thermally driven electrical systems, this means the noise generator must produce voltage fluctuations comparable to $V_T$. This interaction voltage scale applies to a broad class of physical phenomena, including semiconductors and electrochemical reactions \cite{rates}.

However, realizable passive circuits generate weak fluctuations and are not efficient in practical applications. Specifically, in a 1D RC circuit,
\begin{equation}
    \frac{\text{var}(V)^{\text{RC}}}{V_T^2} = \frac{C_T}{C}
\end{equation}
where $C_T = q_e^2/k_b T$ is the thermodynamic capacitance scale, which is approximately $6\unit{\atto\farad}$ at room temperature. Circuits fabricated using the latest transistor processes like the one used here feature capacitances much larger than this, $C \gtrsim 300\unit{\atto\farad}$.

As such, the noise amplification provided by NEAT-RNs is crucial to their utility in room-temperature systems. For the 1D case of our NEAT-RN operating at the balance point of Eqs.~\eqref{eq:g_exp} and \eqref{eq:t_exp}, the variance of the voltage fluctuations is enlarged compared to a passive RC circuit,
\begin{equation}\label{eq:var}
    \text{var}(V) = \frac{1 + \exp\left(V_\text{dd}/2 V_T\right)}{2} \: \text{var}(V)^{\text{RC}}
\end{equation}
the amplification factor depends exponentially on $V_\text{dd}$ and is around $10$ for $V_\text{dd} \approx 6 V_T$.

Because this amplification is intrinsic to the inverter and no complex external circuitry is used, it is achieved efficiently. Specifically, we define the energy consumed by the circuit per random sample as $E=P \tau$, where $\tau$ is the correlation timescale of the output signal, and $P$ is the DC power consumption $P = V_\text{dd} I^*$ given the DC current $I^*$. At the balance point, $\tau$ is 
\begin{equation}
\tau = \frac{C}{G},
\label{eq:tau}
\end{equation}
where $G$ is given by Eq.~\eqref{eq:g_exp}. The energy per sample follows,
\begin{equation}\label{eq:energy_per_sample}
    E = \frac{V_\text{dd} C^2}{q_e} \: \text{var}(V) \: \text{tanh}\left( \frac{V_\text{dd}}{4 V_T}\right)
\end{equation}
where $\text{var}(V)$ is as in Eq.~\eqref{eq:var}.

Eqs.~\eqref{eq:var} and \eqref{eq:tau} elucidate a practically useful property of our system: the timescale and variance of the noise produced by our circuit are independently controllable. $\tau$ can be made arbitrarily large or small (within practical limits) by appropriately manipulating the gate control voltages with respect to $V_\text{dd}$. In contrast, the variance depends only on the powering voltage. From Eq.~\eqref{eq:energy_per_sample}, we can see that the energy consumed per sample does not depend on $\tau$ and scales linearly with the desired variance (in the limit $V_\text{dd} \gg V_T$).

The parameters found from the previous fitting can be used along with Eqs.~\eqref{eq:tau} and \eqref{eq:energy_per_sample} to find that NEAT-RNs could be used as a performant entropy source in a probabilistic computing system. Namely, taking $C \approx 1000 \unit{\atto\farad}$, $I_0^p \approx 1 \unit{\nano\ampere}$, and $\Delta V = 5.5V_T$, we find that $E \approx 15 \unit{\atto\joule}$. In the same scenario, the practical minimum value of $\tau$ will be achieved when $V_p = 0$, at which point $\tau \approx 1  \unit{\nano\second}$.

Overall, we have shown that predictable, programmable, and performant p-circuits can be built using networks of subthreshold transistors. These subthreshold networks can now be easily integrated with other circuitry to build probabilistic computers using advanced transistor processes. We have already proposed one such architecture in \cite{jelincic2025} and hope to report on its implementation in a future contribution.

Moreover, our results motivate a deeper exploration into \textit{probabilistic integrated circuit design}. This burgeoning subfield of analog design is mostly unexplored, offering a rich landscape for new, useful circuit topologies and scientific discoveries. These discoveries could include new theoretical developments, building on the work in \cite{freitas2021stochastic}, or further experimental work to validate these theories.

\section*{Acknowledgements}

TM would like to thank Isaac Chuang for his invaluable advice on experimental design and the writing of the manuscript.

\appendix

\section{Fabrication, control, and measurement of the devices}
\label{ap:device_details}
The experiments conducted in this work utilized a test chip specifically designed to characterize noisy subthreshold circuits in detail. Our chip featured 8-bit DACs with a dynamic range of $[0, 0.175\unit{\volt}]$ that were used for manipulating control voltages. The output signals of each experiment were measured using high-bandwidth and high-input-impedance amplifiers, which were also implemented on the same die. This amplification chain allowed the analog signals to be routed off-chip for measurement using a $1 \unit{\giga\hertz}$ oscilloscope.

\section{Shot-noise models and their diffusive limit}
\label{ap:diffusive_limit}
We consider the stochastic description of non-linear electronic circuits developed in \cite{freitas2021stochastic}, where each conduction device in the circuit exhibits shot noise. The state of the circuit is described by the net number of elementary charges in the free nodes (i.e., the nodes that are not regulated by voltage sources), given by a vector $n \in \mathbb{N}^d$, where $d$ is the number of free nodes. Pairs of free nodes can be connected via conduction devices (diodes, tunnel junctions, transistors, etc.) that are identified by an integer index $\rho >0$. If a conduction device is connected between two free nodes, then elementary charges can jump between them in both directions. 
For each device $\rho$, we assign a transition rate $\lambda_{+\rho}({n})$ to forward conduction events ${n} \to {n} + {\Delta}_\rho$, and a transition rate $\lambda_{-\rho}({n})$ to backward conduction events ${n} \to {n} + {\Delta}_{-\rho}$, with ${\Delta}_{\rho} =-{\Delta}_{-\rho}$. The forward direction is arbitrary, and the vectors $\Delta_\rho$ encode the change in the state $n$ corresponding to each jump or transition. 

For any state $n$, the voltages of the free nodes can be computed as ${v} = q_e C^{-1} {n} + {v}_r$, where $C$ is the Maxwell capacitance matrix of the free nodes, $q_e$ is the charge of the elementary charges, and ${v}_r$ is a constant vector that depends on the regulated voltages. Let $P_t({v})$ be the probability to observe voltages ${v}$ at time $t$. Given the previous description, $P_t({v})$ evolves according to the master equation:
\begin{align}
    \partial_t P_t({v}) &= 
    \sum_{\rho} \lambda_\rho({v} -q_e C^{-1} {\Delta}_\rho) P_t({v} - q_e C^{-1} {\Delta}_\rho) \nonumber \\
    &- \sum_\rho \lambda_\rho({v})P_t({v}), \label{eq:ME}
\end{align}
where we abuse notation by considering the rates $\lambda_\rho({v})$ to be now functions of the voltages ${v}$. A diffusive approximation of this Markov jump process can be obtained by a second-order truncation of the Kramers-Moyal expansion of Eq. \eqref{eq:ME}, which corresponds to the limit of large capacitances $||C|| \gg C_T$. In that case, Eq. \eqref{eq:ME} reduces to a Fokker-Planck equation, which implies that the dynamics of the system can be approximately described by an It\^o stochastic differential equation (SDE) of the form:
\begin{equation}
    C \cdot d{v} =
    {\mu}({v}) dt + \sqrt{2{K}({v})}\cdot d{W}, 
    \label{eq:full_sde}
\end{equation}
where ${W}$ is a vector of independent Wiener processes. This approximation is uncontrolled and fails to capture large fluctuations \cite{hanggi1984bistable, roberts2025}, but properly describes the first and second moments of $P_t({v})$ in the limit of large capacitances \cite{gopal2022large}.
The drift vector ${\mu}({v})$ and diffusion matrix ${K}({v})$ can be obtained from the transition rates $\lambda_{\rho}({v})$. In turn, the two transition rates $\lambda_{\pm \rho}({v})$ associated to the conduction device $\rho$ can be related to its phenomenological IV curve $I_\rho(\Delta v)$ 
via the thermodynamic consistency relations \cite{freitas2021stochastic}:
\begin{equation}
\begin{split}
    \lambda_\rho({v}) - \lambda_{-\rho}({v}) &= I_\rho(\Delta v_\rho)/q_e\\
    \lambda_\rho({v}) + \lambda_{-\rho}({v}) &= \coth\left(\Delta v_\rho /2V_T\right) I_\rho(\Delta v_\rho)/q_e ,
    \label{eq:rates_IV}
\end{split}
\end{equation}
Using Eqs. \eqref{eq:rates_IV}, it is possible to obtain the following expressions for the drift vector and diffusion matrix:
\begin{equation}
\begin{split}
    {\mu}({v}) &\equiv \sum_{\rho} I_{\rho}(\Delta v_\rho) 
    {\Delta}_\rho,\\
    {K}({v}) &\equiv \frac{q_e}{2} \sum_{\rho} I_{\rho}(\Delta v_\rho) \coth\left(\Delta v_\rho/2V_T\right)
    {\Delta}_\rho \cdot {\Delta}_\rho^T.
    \label{eq:drift_diff}
\end{split}
\end{equation}
where $\Delta v_\rho$ is the voltage drop across device $\rho$ in state ${v}$. 

For circuits with deterministic fixed point attractors $v^*$, for which $\mu(v^*) =0$, the stochastic dynamics in Eq. \eqref{eq:full_sde} can be linearized to
\begin{equation}
    C\cdot dv = -G\cdot (v-v^*) \: dt + \sqrt{2k_b G \mathcal{T}} \cdot dW,
    \label{eq:linear_SDE}
\end{equation}
where we have defined the effective conductance matrix $G$ with elements $G_{jk} \equiv -\partial_{v_k} \mu_j(v^*)$ and the effective temperature matrix $\mathcal{T} \equiv G^{-1}K(v^*)/k_b$. The steady-state distribution of the previous dynamics is given by Eqs. \eqref{eq:gaussian_ss} and \eqref{eq:lyapunov} in the main text.

\textcolor{\revcolor}{
Finally, we note that for a one-dimensional case the linearized dynamics of Eq. \eqref{eq:linear_SDE} has the following steady-state autocorrelation function:
\begin{equation}
    \mathbb{X}(\delta t) = \frac{k_b \mathcal{T}}{C} e^{-(G/C)|\delta t|},
\end{equation}
which leads to the following Lorentzian one-sided power spectral density (PSD):
\begin{equation}
    P(f) = \frac{P_0}{1+(f/f_0)^2)},
    \label{eq:lorentz_spectrum}
\end{equation}
with $P_0 = 4k_b\mathcal{T}/G$ and $f_0 = G/(2\pi C)$.
}

\section{EKV-based model of an inverter}
\label{ap:ekv_model}

\textcolor{\revcolor}{
We now consider a single inverter (the left or right pair of transistors in Fig. \ref{fig:main_figure}-(a)). According to the extension of the EKV model including DIBL effects as presented in \cite{low1999cadence}, the current through the NMOS transistor is given by:
\begin{equation}
\begin{split}
    I_n &= I_0^n\log^2\left(1 + e^{(V_{gb} - V_{th}^n)/2n_n} e^{-V_{sb}/2} 
    e^{\gamma_n V_{ds}/2}\right) \\ 
    &- I_0^n\log^2\left(1+ e^{(V_{gb} - V_{th}^n)/2n_n} e^{-V_{db}/2} 
    e^{-\gamma_n V_{ds}/2}\right),
\end{split}
\label{eq:EKV_DIBL}
\end{equation}
where $I_0^n$, $V_{th}^n$, $n_n$ and $\gamma_n$ are model parameters and the voltages $V_{gb} = V_n$, $V_{sb}=0$, $V_{db}=V_{ds}=v$ are here expressed in units of the thermal voltage $V_T$. The same expression gives the current $I_p$ through the PMOS transistor, this time in terms of parameters $I_0^p$, $V_{th}^p$, $n_p$ and $\gamma_p$, by just replacing $I_p \to -I_n$ and using the voltages $V_{gb} = -(V_p-V_{dd})$, $V_{sb}=0$, $V_{db}=V_{ds}=-(v-V_\text{dd})$.
}

It is useful to take a number of approximations in order to make analytical progress. In particular, we consider the subthreshold regime where $e^{(V_{gb}-V_T)/2n_{n/p}} \ll 1$, and neglect DIBL effects (i.e., we set $\gamma_n = \gamma_p =0$). Then, the currents $I_n$ and $I_p$ reduce to the following functions of the output node voltage $v$:
\begin{equation}
\begin{split}
    I_p(v) &= \underbrace{I_0^p e^{-(V_p - V_\text{dd})/n_p}}_{c_p} (1-e^{v-V_\text{dd}}) \\
    I_n(v) &= \underbrace{I_0^n e^{V_n/n_n}}_{c_n}(1-e^{-v}),
\end{split}
\end{equation}
where we have now incorporated factors $e^{-V_{th}^{n/p}/n_{n/p}}$ into the respective constants $I_0^{n/p}$. The deterministic output voltage $v^*$ satisfies $I_p(v^*) = I_n(v^*)$, which under the above approximations leads to the expression:
\begin{equation}
\begin{split}
    e^{v^*} = \frac{e^{V_\text{dd}}}{2} &\left[\!\left(1\!-\!\frac{c_n}{c_p}\right) 
    \!+\! \sqrt{\left(1\!-\!\frac{c_n}{c_p}\right)^2 \!\!+ 4\frac{c_n}{c_p}e^{-V_\text{dd}} } \right] 
    \label{eq:model_fp}
\end{split}
\end{equation}
Using this, we obtain that   the effective conductance is given by:
\begin{equation}
G =V_T^{-1} \sqrt{(c_p-c_n)^2 +4 c_p c_n e^{-V_\text{dd}}},
\end{equation}
Also, the effective temperature is given by:
\begin{equation}
    k_b\mathcal{T} = \frac{q_e I^*}{2G} \left[\coth\left((V_\text{dd}-v^*)/2\right) + \coth\left(v^*/2\right)\right],
\end{equation}
where $I^* \equiv (c_p + c_n - V_T G)/2$ is the stationary current.
The balanced conditions considered in the main text are achieved when $c_n = c_p$. The expressions in the main text also assume that the subthreshold slopes are 1 for all transistors (i.e, $n_{n/p}=1$ for all inverters).

\section{Spectral analysis}
\label{ap:spectral_analysis}

\begin{figure}
    \centering
    \includegraphics[width=\linewidth]{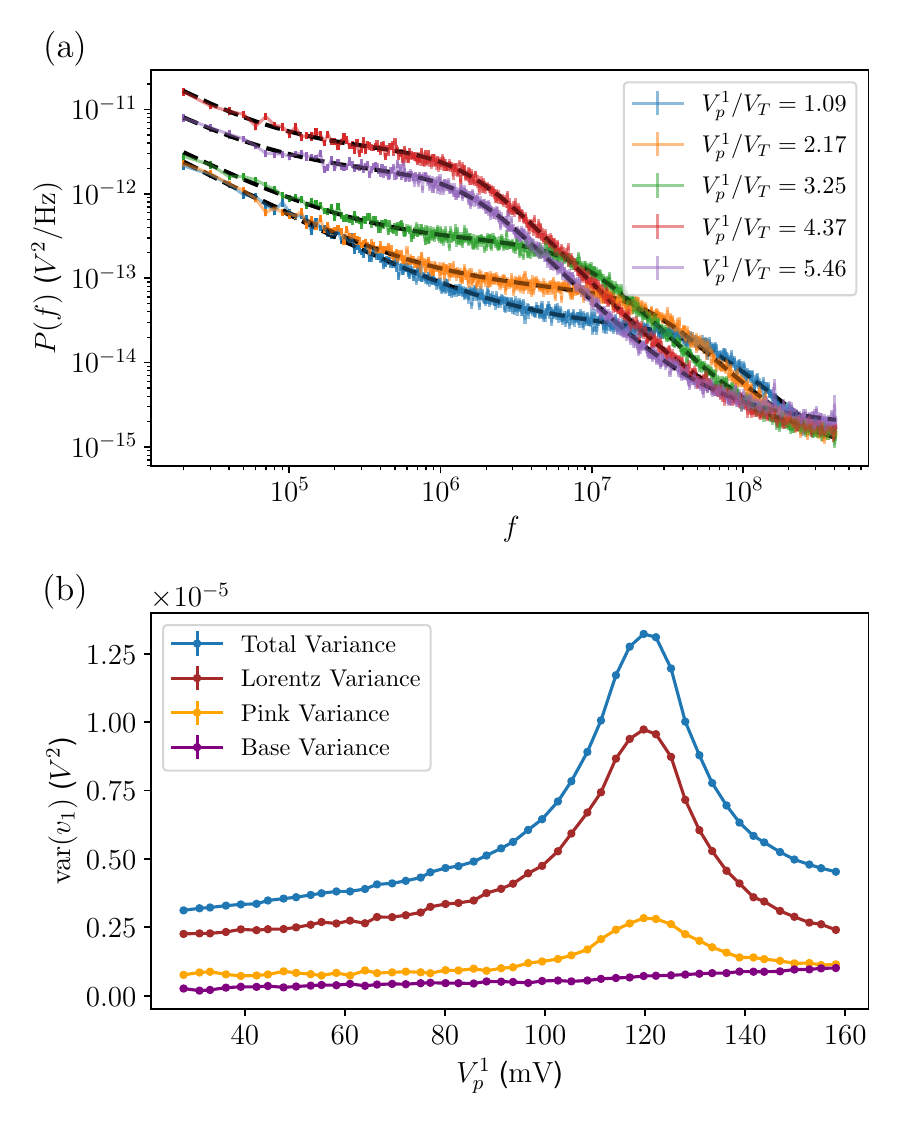}
    \caption{\textcolor{\revcolor}{(a) Power spectral density of the signal $v_1$ for different values of the control voltage $V_p^1$. (b) Total variance of the signal $v_1$, as well as its different spectral contributions, as a function of $V_p^1$. Same settings as in Figure \ref{fig:main_figure}. }}
    \label{fig:spectrum_contributions}
\end{figure}

\textcolor{\revcolor}{
Figure \ref{fig:spectrum_contributions}-(a) shows the power spectral density (PSD) of the signal $v_1$, obtained numerically from the time series, at different operating voltages. The dashed black lines over each PSD show the result of fitting them with the following model:
\begin{equation}
    P(f) = \frac{P_0}{1+(f/f_0)^2} + \frac{L}{f^\alpha} + B.
\end{equation}
The first term corresponds to the Lorentzian spectrum which is expected from the theory in the previous sections. The second term aims to capture the pink noise dominating at low frequencies, while the last term accounts for a flat spectral contribution possibly associated to extrinsic noise sources (for example the amplification stage). This decomposition allows to split the total power or variance of the signal into different contributions, as shown in Fig. \ref{fig:spectrum_contributions}-(b). Finally, we note that the effective temperature can be computed from the parameters $P_0$ and $f_0$ as $k_b\mathcal{T} = Cf_0P_0\pi/2$, where $C$ is the total output capacitance of the inverter in question (for example, for the left inverter it is $C=C_1 + 1/(1/C_2 + 1/C_m)$, in terms of the self-capacitances $C_{1/2}$ and the mutual capacitance $C_m$).
}

\section{Data analysis and model fit}
\label{ap:data_analysis}

\textcolor{\revcolor}{
The comparison of the experimental data with the theoretical model proceeds in three stages: i) determination of intrinsic parameters for each individual inverter, ii) effective modeling of 1/f noise sources and iii) determination of global capacitance matrix based on observed correlations. The raw time series and the code used for the analysis can be found in \cite{dataset}.}

\begin{figure}
    \centering
    \includegraphics[width=\linewidth]{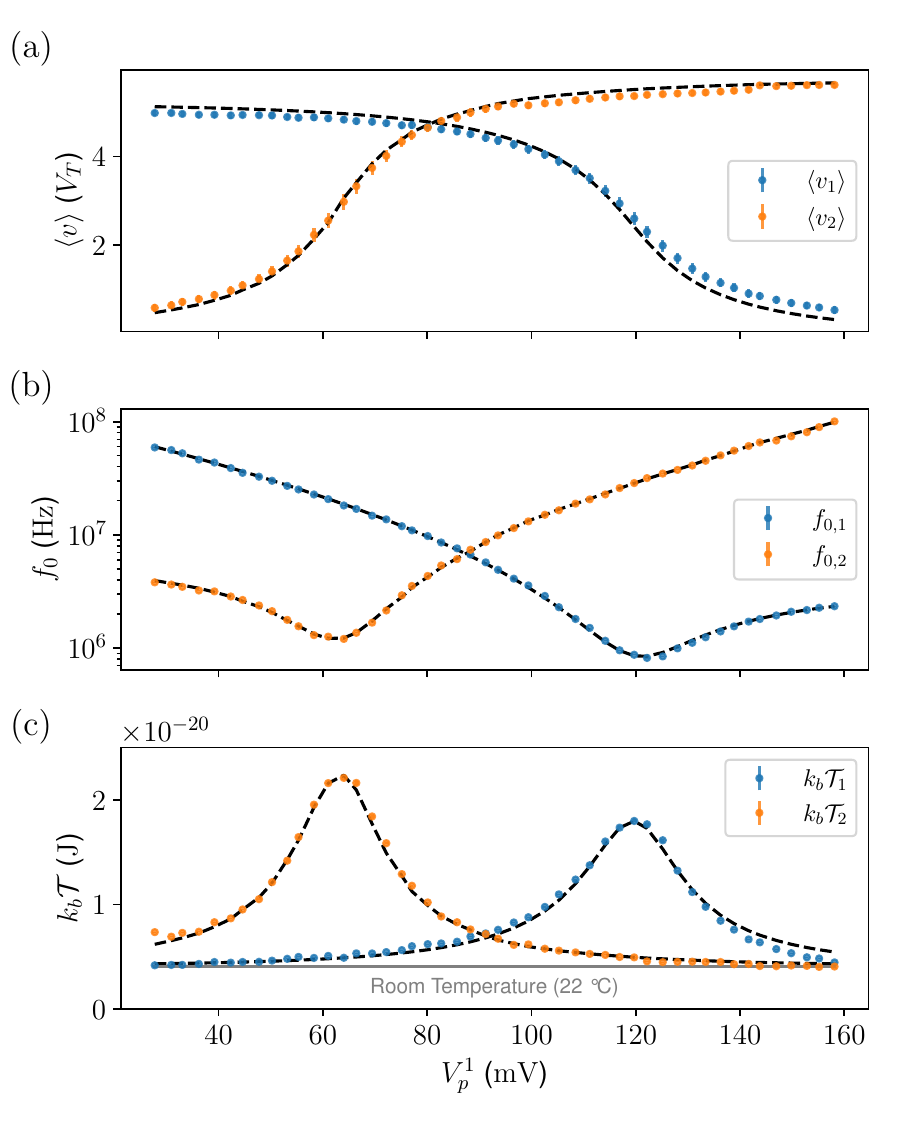}
    \caption{For each individual inverter, (a) average voltage output, (b) frequency $f_0$ and (c) effective temperature $\mathcal{T}$ as obtained from the spectrum (see Eq. \eqref{eq:lorentz_spectrum}). The dashed lines indicate the theoretical results from Eqs. \eqref{eq:drift_diff}, \eqref{eq:linear_SDE} and \eqref{eq:EKV_DIBL} after fitting the three quantities simultaneously. Same settings as in Figure \ref{fig:main_figure}.}
    \label{fig:inverter_fit}
\end{figure}

\textcolor{\revcolor}{In the first stage, for each inverter we simultaneously fit the average output voltage $\mean{v}$, the frequency $f_0$ and the effective temperature $k_b \mathcal{T}$ based on Eqs. \eqref{eq:drift_diff}, \eqref{eq:linear_SDE} and \eqref{eq:EKV_DIBL}. The results are shown in Figure \ref{fig:inverter_fit}. Panels (a) and (b) show that the dynamics of the system is well captured by the model based on Eq. \eqref{eq:EKV_DIBL}, while panel (c) shows that the shot-noise modeling assumption leading to $K(v)$ in Eq. \eqref{eq:drift_diff} is able to describe the intrinsic thermal fluctuations for the operating voltages employed.}

\begin{figure}
    \centering
    \includegraphics[width=\linewidth]{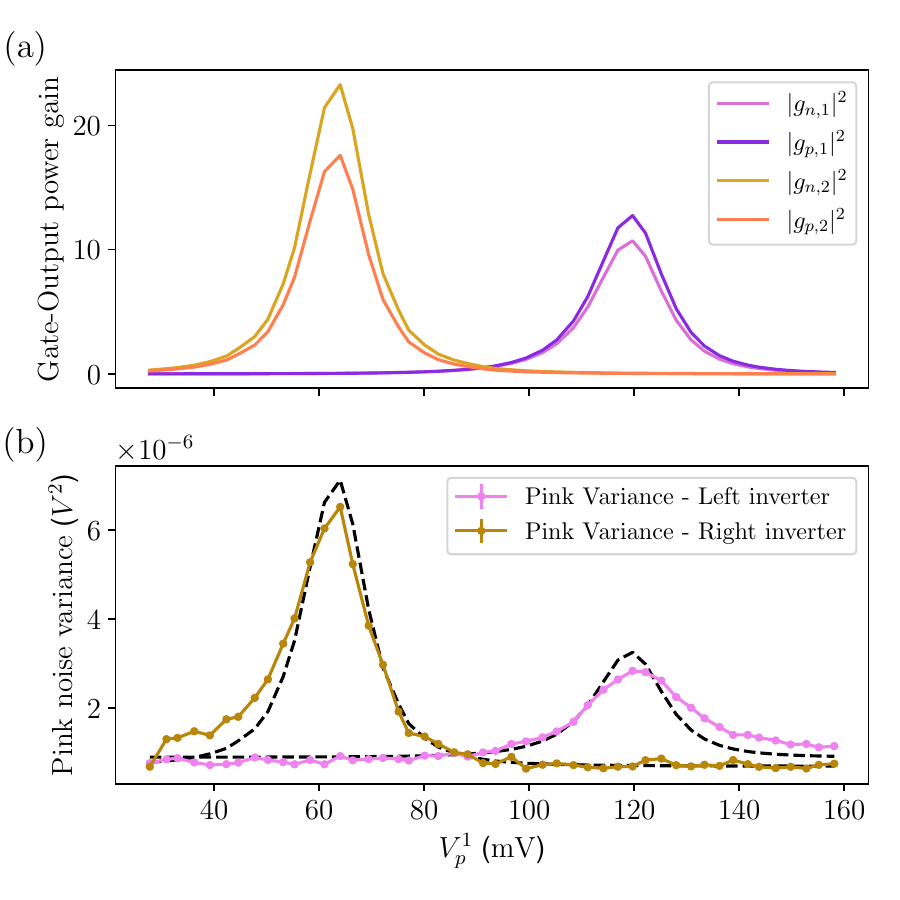}
    \caption{(a) Power gains $|g_{n/p}|^2$ for each inverter, computed from the model with the parameters obtained from the fit in Fig. \ref{fig:inverter_fit}. (b) Total variance of the pink noise for each inverter. Dashed lines corresponds to the best fits according to the model in Eq. \eqref{eq:model_pink_noise}. Same settings as in Figure \ref{fig:main_figure}.}
    \label{fig:pink_noise_fit}
\end{figure}

\textcolor{\revcolor}{
In the second stage we aim to describe the variability of the low-frequency pink noise variance (see Fig. \ref{fig:spectrum_contributions}-(b)), which is the second most important contribution to the total variance in the signal. Low-frequency noise in MOS transistors is usually modeled as stochastic fluctuations in the threshold voltage $V_\text{th}$, caused by the presence of fluctuating charge traps in the gate oxide, which collectively give rise to a 1/f-like spectrum \cite{mahmutoglu2013modeling,stampfer2020advanced,tian2002analysis,jakobson19981,kirton1989noise}. A small-signal analysis of the DC model of each inverter based on Eq. \eqref{eq:EKV_DIBL} offers a simple approach to understand how fluctuations in $V_\text{th}$ are mapped to the measured output. For this, we just need to consider the gains $g_{n/p} = |\partial v^*/\partial V_\text{th}^{n/p}|$, indicating how the deterministic output voltage $v^*$ changes with small perturbations of each of the two threshold voltages. These gains can be easily computed from the model using the parameters fitted in the previous stage (see Fig. \ref{fig:pink_noise_fit}-(a)). Then, if $S_{n/p}$ is the total power of the fluctuations of the threshold voltage $V_\text{th}^{n/p}$ in the frequency range corresponding to the total observation window and sampling rate, the total power of low-frequency fluctuations in the output signal in the same frequency range is given by:
\begin{equation}
    S_{1/f} =  g_n^2 \: S_{n} + g_p^2 \: S_{p} + S_0,
    \label{eq:model_pink_noise}
\end{equation}
where $S_0$ accounts for external sources of low-frequency noise, and it was assumed that the three sources of noise involved are independent. Fig. \ref{fig:pink_noise_fit} compares the total variance due to low-frequency noise as a function of the operating point, as well as the result from the fit using the model in Eq. \ref{eq:model_pink_noise}. We see that, despite its simplicity, this approach allows to explain most of the variability. 
}

\begin{figure}
    \centering
    \includegraphics[width=\linewidth]{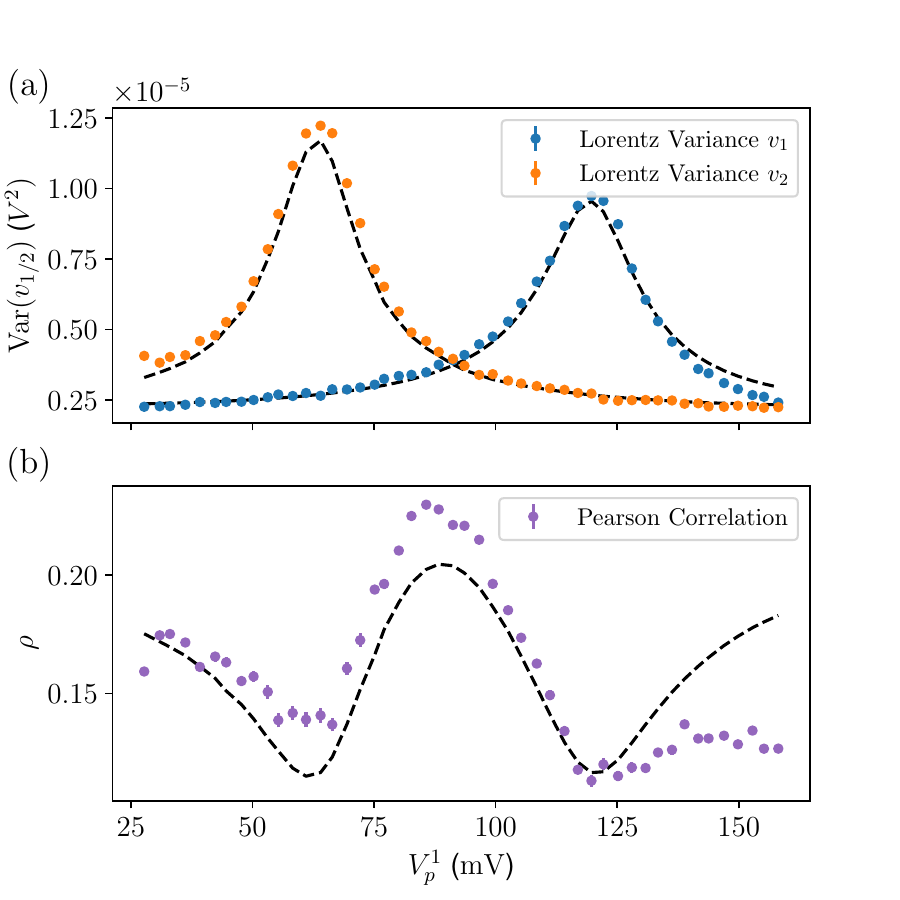}
    \caption{(a) Lorentz variance and (b) Pearson correlation coefficient of the signals $v_{1/2}$ as a function of the operating point. Same settings as in Figure \ref{fig:main_figure}.}
    \label{fig:fit_covariance}
\end{figure}

\textcolor{\revcolor}{
In the last stage we compare the covariance matrix obtained by solving Eq. \eqref{eq:lyapunov} in the main text with the experimentally observed covariance matrix for different operating points, which allows to estimate the self and mutual capacitances of the inverter outputs. This is done as follows. We compute the conductance $G$ and the effective temperature $\mathcal{T}$ for each inverter at each operating point, using the model parameters obtained in the first stage. Then, given a capacitance matrix $C$ constructed from proposed values of $C_1$, $C_2$ and $C_m$, we solve the Lyapunov equation in Eq. \eqref{eq:lyapunov} to obtain an expected covariance matrix $\Sigma$. This covariance matrix is compared to the one estimated from the experimental data, which has the Lorentzian component of the variances as the diagonal elements and the observed correlation $\mathbb{X}_{1,2}(\delta t=0)$ as the non-diagonal elements. The results from this procedure are shown in Fig. \ref{fig:fit_covariance}.
}

\textcolor{\revcolor}{
Finally, Figs. \ref{fig:main_figure}-(d) and (e) in the main text show the aspect ratio and angle of the covariance matrix of the raw time series (that is, including all noise contributions). The dashed lines in the same plots were computed from the fitted model by considering a covariance matrix $\Sigma_T = \Sigma_L + \Sigma_{1/f}$, where $\Sigma_L$ is obtained by solving the Lyapunov equation in Eq. \eqref{eq:lyapunov} and $\Sigma_{1/f}$ is a diagonal matrix with elements $S_{1/f,1}$ and $S_{1/f,2}$ computed according to Eq. \eqref{eq:model_pink_noise}. 
}

\section{Gyration Coefficient}
\label{ap:gyration}
We now sketch the derivation of Eq. \eqref{eq:delayed_corr} for the delayed correlations. We first note that, at steady state conditions, the delayed correlations $\mathbb{X}_{j,k}(\delta t)$ can be written as
\begin{align}
\mathbb{X}_{j,k}(\delta t) &\!=\! \!\int\!\! dv \!\!\int\!\! dv' (v'_k \!-\! v^*_k) (v_j-v^*_j) P({v}',t+\delta t| {v}, t) P(v) \nonumber \\
&\!=\! \!\int\! \!dv \: (v_j\!-\!v^*_j) \left [e^{-\delta t \: C^{-1} G}\cdot (v\!-\!v^*)  \right]_k P(v) \nonumber \\
&= \left[e^{-\delta tC^{-1}G} \: \Sigma \right]_{k,j}
= \left[\Sigma \: e^{-\delta t G C^{-1}}\right]_{j,k},
\end{align}
where we have used that, for the linear process in Eq. \eqref{eq:linear_SDE}, the conditional probability $P({v}',t+\delta t| {v}, t)$ is a Gaussian distribution with an exponentially relaxing mean value 
$\langle v'\rangle = v^* + e^{-\delta t \: C^{-1} G}\cdot (v\!-\!v^*)$, and that the stationary distribution $P(v)$ is given by Eq. \eqref{eq:gaussian_ss}. Eq. \eqref{eq:delayed_corr} follows by just using that $\int_0^{t_I} d\tau e^{-\tau A} = A^{-1} (\mathds{1} - e^{-t_I A})$ for any invertible matrix $A$.

\bibliography{references}

\end{document}